\documentclass[epj]{svjour}
\usepackage{graphicx}
\usepackage{dcolumn}
\usepackage{multirow}
\usepackage{amssymb}
\onecolumn
\begin{document}
\title{Isospin Symmetry Breaking and Octet Baryon Masses \\ due to Their Mixing with Decuplet Baryons}

\author{Hua-Xing Chen
}                     
\institute{School of Physics and Nuclear Energy Engineering and International Research Center for Nuclei and Particles in the Cosmos, Beihang University, Beijing 100191, China}
\date{Received: date / Revised version: date}
%
\abstract{We study the isospin symmetry breaking and mass splittings of the eight lowest-lying baryons. We consider three kinds of baryon mass terms, including the bare mass term, the electromagnetic terms and the spontaneous chiral symmetry breaking terms. We include the mixing term between flavor-octet and flavor-decuplet baryons. This assumes that the lowest-lying $\Sigma$ and $\Xi$ baryons contain a few decuplet components and so are not purely flavor-octet. We achieve a good fitting that the difference between every fitted mass and its experimental value is less than 0.2 MeV.
} 
\maketitle
%
%

\section{Introduction}
\label{Intro}

Isospin symmetry breaking in mass splittings of the lowest-lying flavor-octet baryons is of fundamental importance~\cite{Feynman,Gasser:1982ap}. Lots of efforts and many methods have been devoted to study it, such as the chiral perturbation theory ($\chi$PT)~\cite{Shanahan:2012wa,Shanahan:2013xw}, the chiral soliton model~\cite{Praszalowicz:1992gn,Yang:2010id}, the Cottingham's sum rule~\cite{WalkerLoud:2012bg}, the QCD sum rule~\cite{Adami:1993xz,Forkel:1996ty,Henley:1996hr,Narison:2002pw,ioffe}, the Dashen's theorem~\cite{dashen,Donoghue:1993hj,Bijnens:1993ae} and the Skyrme Model~\cite{Li:1986iya,Jain:1989kn}, etc.. Isospin symmetry breaking originates from two different sources: electromagnetic (EM) self-energies and the current quark mass differences. Moreover, due to their influences, the quark condensates are also different, i.e., $\langle \bar u u \rangle \neq \langle \bar d d \rangle \neq \langle \bar s s \rangle$, and so the $SU(3)_L \otimes SU(3)_R$ chiral symmetry and the isospin symmetry are both explicitly and spontaneously broken in a dissymmetric way. In general we can separate baryon masses into hadronic and electromagnetic parts, which are contributed by QCD and electromagnetic effects, respectively. In recent years, lattice QCD is developing very fast and several lattice groups have determined these two contributions~\cite{Durr:2008zz,Duncan:2004ys,Basak:2013iw,Shanahan:2012wa,Bhattacharya:2013ehc,Green:2012ej,Beane:2006fk,Horsley:2012fw,Borsanyi:2013lga,deDivitiis:2013xla,Blum:2010ym,Bietenholz:2011qq,Aoki:2008sm} (reviewed in Refs.~\cite{Colangelo:2010et,Fodor:2012gf,Portelli:2013jla}).

In this paper we shall study the isospin symmetry breaking and mass splittings of the eight lowest-lying baryons. We assume that their masses originate from three different sources: the bare mass term, the electromagnetic terms and the spontaneous chiral symmetry breaking terms. Particularly, we shall use group theoretical methods to study the spontaneous chiral symmetry breaking terms~\cite{Weinberg:1969hw,Weinberg:1990xn,Leinweber:1994nm,Ioffe:1981kw,Chung:1981cc,Espriu:1983hu,Cohen:1996sb,Nagata:2007di,Cohen:2002st,Jido:2001nt,Chen:2012ut,Chen:2013gnu,Chen:2013jra}.
Using this method we have studied local and non-local three-quark baryon fields, and found there are five chiral multiplets~\cite{Chen:2008qv,Dmitrasinovic:2011yf,Chen:2012vs,Chen:2013efa}:
\begin{eqnarray}
[(\mathbf{3}, \mathbf{1})\oplus(\mathbf{1}, \mathbf{3})]^3 &=& [(\mathbf{1}, \mathbf{1})] \oplus [(\mathbf{8}, \mathbf{1})\oplus(\mathbf{1}, \mathbf{8})] \oplus [(\mathbf{10}, \mathbf{1})\oplus(\mathbf{1}, \mathbf{10})]
\\ \nonumber && \oplus [(\mathbf{\bar 3}, \mathbf{3})\oplus(\mathbf{3}, \mathbf{\bar 3})] \oplus [(\mathbf{6}, \mathbf{3})\oplus(\mathbf{3}, \mathbf{6})] \, ,
\end{eqnarray}
as well as their mirror partners. Because the chiral symmetry is broken, these five multiplets do not actually exist. They mix and compose flavor-singlet, flavor-octet and flavor-decuplet baryons, and then compose our real world:
\begin{eqnarray}
\nonumber [(\mathbf{1}, \mathbf{1})] \& [(\mathbf{\bar 3}, \mathbf{3})\oplus(\mathbf{3}, \mathbf{\bar 3})] &\rightarrow& \mathbf{1}_F \, ,
\\ \nonumber [(\mathbf{8}, \mathbf{1})\oplus(\mathbf{1}, \mathbf{8})] \& [(\mathbf{\bar 3}, \mathbf{3})\oplus(\mathbf{3}, \mathbf{\bar 3})] \& \underline{[(\mathbf{6}, \mathbf{3})\oplus(\mathbf{3}, \mathbf{6})]} &\rightarrow& \mathbf{8}_F \, ,
\\ \nonumber [(\mathbf{10}, \mathbf{1})\oplus(\mathbf{1}, \mathbf{10})] \& \underline{[(\mathbf{6}, \mathbf{3})\oplus(\mathbf{3}, \mathbf{6})]} &\rightarrow& \mathbf{10}_F \, .
\end{eqnarray}
Usually the eight lowest-lying baryons are considered to be ``purely'' flavor-octet. However, it might be possible that flavor-octet and flavor-decuplet baryons can mix through their common $[(\mathbf{6}, \mathbf{3})\oplus(\mathbf{3}, \mathbf{6})]$ chiral components, and so they are not pure any more. For example, the lowest-lying $\Sigma(1193)$ baryons can mix with the higher $\Sigma$ baryons having the same quantum numbers $J^P = 1/2^+$. Several candidates are $\Sigma(1660)$, $\Sigma(1770)$ and $\Sigma(1880)$, which may contain flavor-decuplet components. We note that the $[(\mathbf{6}, \mathbf{3})\oplus(\mathbf{3}, \mathbf{6})]$ chiral multiplet is important to explain the experimental value of the axial charge $g_A = 1.267$~\cite{Beringer:1900zz,Chen:2009sf}.

In this paper we assume the mixing of flavor-octet and flavor-decuplet baryons can contribute to flavor-octet baryon masses. To study this effect, we shall use the chiral invariant Lagrangians $\bar B M B$ obtained in Refs.~\cite{Chen:2010ba,Chen:2011rh}, where $\bar B$ and $B$ denote baryon fields, and $M$ denotes meson fields. However, these Lagrangians contain too many parameters to be solvable, and so we shall first use them to obtain flavor-singlet Lagrangians and then obtain the spontaneous chiral symmetry breaking mass terms. By assuming this mixing the lowest-lying $\Sigma$ and $\Xi$ baryons contain a few decuplet components and so are not purely flavor-octet. Although these contributions may be quite small, they can still be important, because masses of the lowest-lying flavor-octet baryons have been measured so accurately nowadays. We note that flavor-singlet and flavor-octet baryons can also mix through their common $[(\mathbf{\bar 3}, \mathbf{3})\oplus(\mathbf{3}, \mathbf{\bar 3})]$ chiral components, which we shall not study in this paper~\cite{future}.

This paper is organized as follows. In Sec.~\ref{sec:transform} we transform from chiral-singlet Lagrangians to flavor-singlet ones. In Sec.~\ref{sec:formulae} we introduce the formulae used to calculate the eight lowest-lying baryon masses. We consider three kinds of terms: the bare mass term, the electromagnetic terms and the spontaneous chiral symmetry breaking terms. The spontaneous chiral symmetry breaking terms are discussed in Sec.~\ref{sec:octet} and Sec.~\ref{sec:mixing}, other terms are discussed in Sec.~\ref{sec:others}, and their summation is shown in Sec.~\ref{sec:octetformulae}. In Sec.~\ref{sec:numerical} we use these formulae to perform numerical analyses and do the fitting. Sec.~\ref{sec:summary} is a summary.

\section{From Chiral-Singlet Lagrangians to Flavor-Singlet Lagrangians}
\label{sec:transform}

The chiral invariant Lagrangians $\bar B M B$ have been obtained in Refs.~\cite{Chen:2010ba,Chen:2011rh}, where $\bar B$ and $B$ denote baryon fields, and $M$ denotes meson fields. However, they contain too many parameters to be solvable. Therefore, in this section we transform from these chiral-singlet Lagrangians to flavor-singlet ones, which will be used to calculate the spontaneous chiral symmetry breaking terms in the next section.

In Refs.~\cite{Chen:2008qv,Dmitrasinovic:2011yf,Chen:2012vs,Chen:2013efa} we use group theoretical methods to study local and non-local three-quark baryon fields, and found there are five chiral multiplets: $[(\mathbf{1}, \mathbf{1})]$, $[(\mathbf{8}, \mathbf{1})\oplus(\mathbf{1}, \mathbf{8})]$, $[(\mathbf{10}, \mathbf{1})\oplus(\mathbf{1}, \mathbf{10})]$, $[(\mathbf{\bar 3}, \mathbf{3})\oplus(\mathbf{3}, \mathbf{\bar 3})]$ and $[(\mathbf{6}, \mathbf{3})\oplus(\mathbf{3}, \mathbf{6})]$ as well as their mirror partners. We note that only non-local three-quark baryon fields can belong to the $[(\mathbf{1}, \mathbf{1})]$ chiral multiplet~\cite{Dmitrasinovic:2011yf,Chen:2012vs,Chen:2013efa}. Because the chiral symmetry is broken, these five multiplets do not actually exist. These chiral multiplets mix and compose the physical flavor-singlet ($\Lambda$), flavor-octet ($N$) and flavor-decuplet ($\Delta$) baryons, and then compose our real world:
\begin{eqnarray}
\nonumber | \Lambda \rangle &=& \alpha_1 | \Lambda_{[(\mathbf{1}, \mathbf{1})]} \rangle + \alpha_2 | \Lambda_{[(\mathbf{\bar 3}, \mathbf{3})]} \rangle \, ,
\\ | N \rangle &=& \beta_1 | N_{[(\mathbf{8}, \mathbf{1})]} \rangle + \beta_2 | N_{[(\mathbf{\bar 3}, \mathbf{3})]} \rangle + \beta_3 | N_{[(\mathbf{6}, \mathbf{3})]} \rangle \, ,
\\ \nonumber | \Delta \rangle &=& \gamma_1 | \Delta_{[(\mathbf{10}, \mathbf{1})]} \rangle + \gamma_2 | \Delta_{[(\mathbf{6}, \mathbf{3})]} \rangle \, .
\end{eqnarray}

We use the $[(\mathbf{6}, \mathbf{3})\oplus(\mathbf{3}, \mathbf{6})]$ chiral multiplet as an example to show how to obtain flavor-singlet Lagrangians from chiral-singlet ones. The chiral-singlet Lagrangian has been obtained in  Ref.~\cite{Chen:2010ba}:
\begin{equation}
g_{(18)} \bar N_{(18)}^a (\sigma^c + i \gamma_5 \pi^c) ({\bf D}_{(18)}^c)_{ab} N_{(18)}^b \, ,
\label{def:18 int}
\end{equation}
where $a,b=1\cdots18$, $c=0\cdots8$, $g_{(18)}$ is the coupling constant and $\sigma^c + i \gamma_5 \pi^c$ are meson fields belonging to the $[(\mathbf{\bar 3}, \mathbf{3})\oplus(\mathbf{3}, \mathbf{\bar 3})]$ chiral representation. The baryon field $N^a_{(18)}$ belongs to the chiral representation $[({\bf 6},{\bf 3})\oplus({\bf 3},{\bf 6})]$. It contains both flavor-octet and flavor-decuplet baryons:
\begin{eqnarray}
N^a_{(18)} &=& (N_\mu^N, \Delta_\mu^P)^T = (N_{[(\mathbf{6}, \mathbf{3})]}^N, \Delta_{[(\mathbf{6}, \mathbf{3})]}^P)^{\rm T} \, ,
\end{eqnarray}
where $N=1\cdots8$ and $P=1\cdots10$; $\epsilon_{abc}$ and $S_P^{ABC}$ are the totally anti-symmetric tensor and totally symmetric tensor, respectively; $N_\mu^N$ and $\Delta_\mu^P$ are the octet and decuplet baryons belonging to the $[(\mathbf{6}, \mathbf{3})\oplus(\mathbf{3}, \mathbf{6})]$ chiral multiplet~\cite{Chen:2008qv}. The matrices ${\bf D}_{(18)}$ are:
\begin{eqnarray}
{\bf D}^0_{(18)} &=& {1\over\sqrt6}
\left ( \begin{array}{cc}
{\bf 1}_{8\times8} & 0 \\
0 & -2 \times {\bf 1}_{10\times10}
\end{array} \right ) \, ,
\label{def:D18}
\\ \nonumber {\bf D}^a_{(18)} &=&
\left ( \begin{array}{cc}
{\rm {\bf D}_{(8)}^{a} + {2\over3} {\bf F}_{(8)}^{a}} & -{1\over\sqrt3}{\rm {\bf T}_{(8/10)}^a }
\\ -{1\over\sqrt3}{\rm {\bf T}^{\dagger a}_{(8/10)} } & -{2\over3} {\bf F}_{(10)}^{a}
\end{array} \right ) \, ,
\end{eqnarray}
where ${\bf D}_{(8)}^a$, ${\bf F}_{(8)}^a$, ${\rm {\bf T}}_{(8/10)}^a$ and ${\bf F}_{(10)}^{a}$ are listed in Ref.~\cite{Chen:2009sf}. From this chiral-singlet Lagrangian we obtain five flavor-singlet Lagrangians:
\begin{eqnarray}
\nonumber && g_{(18)} \bar N_{[(\mathbf{6}, \mathbf{3})]}^N \left( (\sigma^0 + i \gamma_5 \pi^0) ({1\over\sqrt6}\delta_{NM}) \right) N_{[(\mathbf{6}, \mathbf{3})]}^M \, ,
\\ \nonumber && g_{(18)} \bar N_{[(\mathbf{6}, \mathbf{3})]}^N \left( (\sigma^c + i \gamma_5 \pi^c) ({\rm {\bf D}_{(8)}^{c} + {2\over3} {\bf F}_{(8)}^{c}})_{NM} \right) N_{[(\mathbf{6}, \mathbf{3})]}^M \, ,
\\ && g_{(18)} \bar \Delta_{[(\mathbf{6}, \mathbf{3})]}^P \left( (\sigma^0 + i \gamma_5 \pi^0) (-{2\over\sqrt6} \delta_{PQ}) \right) \Delta_{[(\mathbf{6}, \mathbf{3})]}^Q \, ,
\\ \nonumber && g_{(18)} \bar \Delta_{[(\mathbf{6}, \mathbf{3})]}^P \left( (\sigma^c + i \gamma_5 \pi^c) (-{2\over3} {\bf F}_{(10)}^{c})_{PQ} \right) \Delta_{[(\mathbf{6}, \mathbf{3})]}^Q \, ,
\\ \nonumber && g_{(18)} \bar N_{[(\mathbf{6}, \mathbf{3})]}^N (\sigma^c + i \gamma_5 \pi^c) (-{1\over\sqrt3}{\rm {\bf T}_{(8/10)}^c })_{NP} \Delta_{[(\mathbf{6}, \mathbf{3})]}^P + h.c. \, .
\end{eqnarray}
where $N,M=1\cdots8$, $P,Q=1\cdots10$ and $c=1\cdots8$. 

It seems that it is much more complicated to use flavor-singlet Lagrangians than chiral-singlet ones, because there are much more flavor-singlet Lagrangians. However, it turns out to be that using these flavor-singlet Lagrangians is much simpler, because the only possible matrices to connect $\bar B M B$ are identity matrices (${\bf 1}_{1\times1}$, ${\bf 1}_{8\times8}$ and ${\bf 1}_{10\times10}$), ${\bf D}_{(8)}^{a}$, ${\bf F}_{(8)}^{a}$, ${\bf F}_{(10)}^{a}$, ${\bf T}^a_{(1/8)}$ and ${\bf T}_{(8/10)}^a$~\cite{Chen:2009sf}.

Summarizing all these flavor-singlet Lagrangians, we obtain six diagonal terms
\begin{eqnarray}
\label{def:11} && g_1 \bar \Lambda \sigma^0 \Lambda \, ,
\\ \label{def:88} && g_2 \bar N^N \sigma^0 N^N \, ,
\\ \label{def:88D} && g_3 \bar N^N \sigma^c ({\bf D}_{(8)}^{c})_{NM} N^M \, ,
\\ \label{def:88F} && g_4 \bar N^N \sigma^c ({\bf F}_{(8)}^{c})_{NM} N^M \, ,
\\ \label{def:10} && g_5 \bar \Delta^P \sigma^0 \Delta^P \, ,
\\ \label{def:1010} && g_6 \bar \Delta^P \sigma^c ({\bf F}_{(10)}^{c})_{PQ} \Delta^Q \, ,
\end{eqnarray}
and two off-diagonal terms
\begin{eqnarray}
\label{def:18} && g_7 \bar \Lambda \sigma^c ({\bf T}^c_{1/8})_{1N} N^N + h.c. \, ,
\\ \label{def:810} && g_8 \bar N^N \sigma^c ({\bf T}_{(8/10)}^c)_{NP} \Delta^P + h.c. \, ,
\end{eqnarray}
where $N,M=1\cdots8$, $P,Q=1\cdots10$ and $c=1\cdots8$. Some readers may be quite familiar with these Lagrangians because they are similar to the lowest-order meson-baryon $\chi$PT Lagrangian in the flavor space. In these equations every flavor-singlet Lagrangian is the summation of several chiral components~\cite{Chen:2009sf,Chen:2010ba}, for example, $\bar N^N \sigma^c ({\bf D}_{(8)}^{c})_{NM} N^M$ is the summation of
\begin{eqnarray}
\bar N^N \sigma^c ({\bf D}_{(8)}^{c})_{NM} N^M &\sim& a_1 \bar N_{[(\mathbf{6}, \mathbf{3})]}^N \sigma^c ({\bf D}_{(8)}^{c})_{NM} N_{[(\mathbf{6}, \mathbf{3})]}^M
\\ \nonumber &+& a_2 \bar N_{[(\mathbf{\bar 3}, \mathbf{3})]}^N \sigma^c ({\bf D}_{(8)}^{c})_{NM} N_{[(\mathbf{\bar 3}, \mathbf{3})]}^M
\\ \nonumber &+& a_3 \bar N_{[(\mathbf{8}, \mathbf{1})]}^N \sigma^c ({\bf D}_{(8)}^{c})_{NM} N_{[(\mathbf{6}, \mathbf{3})]}^M
\\ \nonumber &+& a_4 \bar N_{[(\mathbf{8}, \mathbf{1})]}^N \sigma^c ({\bf D}_{(8)}^{c})_{NM} N_{[(\mathbf{\bar 3}, \mathbf{3})]}^M
\\ \nonumber &+& h.c. + \cdots \, .
\end{eqnarray}
Here we only include flavor-singlet terms containing scalar meson fields $\sigma^a$, which have nonzeron condensates $\langle \sigma_{0,3,8} \rangle$. We note that the Lagrangians containing $\pi^a$ can also contribute, and their contributions can be calculated using many methods, such as the chiral perturbation theory~\cite{Shanahan:2012wa,Shanahan:2013xw,Weinberg:1978kz,Gasser:1983yg,Bernard:1995dp,Pich:1995bw,Ecker:1994gg,Scherer:2012xha,Pascalutsa:1998pw,Pascalutsa:1999zz,Ren:2013oaa}. However, they are beyond the scope of this paper, and we shall omit their contributions.

In this paper we only consider the mixing of flavor-octet and flavor-decuplet baryons, and so Eqs.~(\ref{def:11}) and (\ref{def:18}) are irrelevant. Moreover, we assume all the decuplet baryons have the same mass $m_{\Delta} =2$ GeV, and so we only need Eqs.~(\ref{def:88}), (\ref{def:88D}), (\ref{def:88F}), (\ref{def:10}) and (\ref{def:810}). This is because that the lowest $\Delta$ baryon having quantum numbers $J^P = {1\over2}^+$ is $\Delta(1910)$~\cite{Beringer:1900zz}; besides it, there can be other decuplet baryons having the same quantum numbers $J^P = {1\over2}^+$ but having larger masses, which can also mix with the lowest-lying flavor-octet baryons. The dependence of our results on $m_{\Delta}$ will be studied in Sec.~\ref{sec:decupletmass}.

\section{Mass Formulae}
\label{sec:formulae}

\subsection{Flavor-Octet Baryons}
\label{sec:octet}

Using Lagrangians (\ref{def:88}), (\ref{def:88D}) and (\ref{def:88F}) we can investigate the $SU(3)$ flavor symmetry breaking effects through the terms containing nonzero condensate $\langle \sigma_8 \rangle$, and the $SU(2)$ isospin symmetry breaking effects through the terms containing nonzero condensate $\langle \sigma_3 \rangle$. Their explicit forms after the spontaneous chiral symmetry breaking are
\begin{eqnarray}
\mathcal{L}^{{\rm after}}_{N} &=& g_2 \langle \sigma_0 \rangle \times \Big ( \bar N N + \bar \Sigma \Sigma + \bar \Xi \Xi + \bar \Lambda^0 \Lambda^0 \Big )
\\ \nonumber &+& g_3 \langle \sigma_8 \rangle \times \Big ( - {1\over2\sqrt3} \bar N N + {1\over\sqrt3} \bar \Sigma \Sigma - {1\over2\sqrt3} \bar \Xi \Xi - {1\over\sqrt3} \bar \Lambda^0 \Lambda^0 \Big )
\\ \nonumber &+& g_4 \langle \sigma_8 \rangle \times \Big ({\sqrt3\over2} \bar N N - {\sqrt3\over2} \bar \Xi \Xi \Big )
\\ \nonumber &+& g_3 \langle \sigma_3 \rangle \times \Big ( {1\over2}\bar p^+ p^+ - {1\over2} \bar n^0 n^0 - {1\over2} \bar \Xi^0 \Xi^0 + {1\over2} \bar \Xi^- \Xi^- + {1\over\sqrt3} \bar \Lambda^0 \Sigma^0 + {1\over\sqrt3} \bar \Sigma^0 \Lambda^0 \Big )
\\ \nonumber &+& g_4 \langle \sigma_3 \rangle \times \Big ( {1\over2} \bar p^+ p^+ - {1\over2} \bar n^0 n^0 + \bar \Sigma^+ \Sigma^+ - \bar \Sigma^- \Sigma^- + {1\over2} \bar \Xi^0 \Xi^0 - {1\over2} \bar \Xi^- \Xi^- \Big ) \, ,
\end{eqnarray}
where $N = (p^+, n^0)^{\rm T}$, $\Sigma = (\Sigma^+, \Sigma^0, \Sigma^-)^{\rm T}$ and $\Xi = (\Xi^0, \Xi^-)^{\rm T}$. We note that $\Sigma^0$ and $\Lambda^0$ can mix with each other when $g_3 \langle \sigma_3 \rangle \neq 0$.

\subsection{Mixing of Flavor-Octet and Flavor-Decuplet Baryons}
\label{sec:mixing}

Using the Lagrangian (\ref{def:810}) we can investigate the isospin symmetry breaking effects through mixing terms containing nonzero condensates $\langle \sigma_3 \rangle$ and $\langle \sigma_8 \rangle$. Its explicit form after the spontaneous chiral symmetry breaking is
\begin{eqnarray}
\mathcal{L}^{{\rm after}}_{N\Delta} &=& g_8 \langle \sigma_8 \rangle \times \Big ( {1 \over \sqrt2} \bar \Sigma \Sigma^* - {1 \over \sqrt2} \bar \Xi \Xi^* + {1 \over \sqrt2} \bar \Sigma^* \Sigma - {1 \over \sqrt2} \bar \Xi^* \Xi \Big )
\\ \nonumber &+& g_8 \langle \sigma_3 \rangle \times \Big ( \sqrt {2 \over 3} \bar p^+ \Delta^{+} + \sqrt {2 \over 3} \bar n^0 \Delta^{0} + {1 \over \sqrt6} \bar \Sigma^+ \Sigma^{*+} - {1 \over \sqrt6} \bar \Sigma^- \Sigma^{*-} - {1 \over \sqrt6} \bar \Xi^0 \Xi^{*0} + {1 \over \sqrt6} \bar \Xi^- \Xi^{*-} - {1 \over \sqrt2} \bar \Lambda^0 \Sigma^{*0}
\\ \nonumber && ~~~~~ + \sqrt {2 \over 3} \bar \Delta^{+} p^+ + \sqrt {2 \over 3} \bar \Delta^{0} n^0 + {1 \over \sqrt6} \bar \Sigma^{*+} \Sigma^+ - {1 \over \sqrt6} \bar \Sigma^{*-} \Sigma^- - {1 \over \sqrt6} \bar \Xi^{*0} \Xi^0 + {1 \over \sqrt6} \bar \Xi^{*-} \Xi^- - {1 \over \sqrt2} \bar \Sigma^{*0} \Lambda^0 \Big ) \, ,
\end{eqnarray}
where $\Sigma^* = (\Sigma^{*+}, \Sigma^{*0}, \Sigma^{*-})^{\rm T}$ and $\Xi^* = (\Xi^{*0}, \Xi^{*-})^{\rm T}$. We find that the effects of terms containing $\langle \sigma_3 \rangle$ and $\langle \sigma_8 \rangle$ have some overlaps: the former is used to decrease $\Sigma$ and $\Xi$ masses, while the latter is used to decrease the masses of $p^+$, $n^0$, $\Sigma^+$, $\Sigma^-$, $\Xi^0$ and $\Xi^-$.

\subsection{Other Contributions}
\label{sec:others}

Besides the spontaneous chiral symmetry breaking terms listed in previous subsections, octet baryon masses are contributed by two other sources:
\begin{enumerate}

\item The ``bare'' mass term $m_{bare} \bar N^N N^N$. Numerically it is equivalent to the spontaneous chiral symmetry breaking term $g_2 \langle \sigma_0 \rangle \bar N^N N^N$. We write them together as $m_0 \bar N^N N^N$. It conserves the $SU(3)$ flavor symmetry.

\item The electromagnetic terms, which break the isospin symmetry. We can not calculate these terms directly, but just use two simple schemes to evaluate them. They both contain only one free parameter, which will be used to do fitting in Sec.~\ref{sec:numerical}. 
    The first scheme, ``Scheme A'', simply assumes masses of charged and unchanged baryons are different, for examples, $EM^A_{p^+} = EM^A_{\Sigma^+} = EM_A$ and $EM^A_{n^0} = EM^A_{\Sigma^0} = 0$. The second scheme, ``Scheme B'', simply calculates electromagnetic interactions among the three valence quarks, for example, $EM^B_{p^+} = EM_B \times \big ( {2\over3}{2\over3} - {1\over3}{2\over3} - {1\over3}{2\over3} \big ) = 0$ and $EM^B_{n^0} = EM_B \times \big ( -{2\over3}{1\over3} -{2\over3}{1\over3} + {1\over3}{1\over3} \big ) = -{1\over3} EM_B$. These two assumptions are listed in Tab.~\ref{tab:EM}. We note that the Scheme A considers baryons as a whole, while the Scheme B considers baryons as three-quark objects. They both satisfy the well-known Coleman-Glashow mass formula $\left(M_{p^+} - M_{n^0}\right)_{EM} + \left(M_{\Xi^0} - M_{\Xi^-}\right)_{EM} = \left(M_{\Sigma^+} - M_{\Sigma^-}\right)_{EM}$~\cite{Coleman}.

\item We note that the decuplet baryon masses are all assumed to be $m_{\Delta} =2$ GeV.

\end{enumerate}

\begin{table}[tbh]
\begin{center}
\caption{The electromagnetic terms under the two assumptions, the Scheme A and the Scheme B. They both contain only one free parameter, which will be used to do fitting in Sec.~\ref{sec:numerical}.}
\begin{tabular}{ccccccccc}
\hline \hline
& $p^+$ & $n^0$ & $\Sigma^+$ & $\Sigma^0$ & $\Sigma^-$ & $\Xi^0$ & $\Xi^-$ & $\Lambda^0$
\\ \hline
Scheme A & $EM_A$ & 0 & $EM_A$ & 0 & $EM_A$ & 0 & $EM_A$ & 0
\\ \hline
Scheme B & 0 & $-{1\over3} EM_B$ & 0 & $-{1\over3} EM_B$ & ${1\over3} EM_B$ & $-{1\over3} EM_B$ & ${1\over3} EM_B$ & $-{1\over3} EM_B$
\\ \hline \hline
\end{tabular}
\label{tab:EM}
\end{center}
\end{table}

\subsection{Flavor-Octet Baryon Mass Formulae}
\label{sec:octetformulae}

Summarizing all the possible terms we arrive at the formulae to calculate octet baryon masses. There are diagonal and off-diagonal terms.

The diagonal terms are:
\begin{eqnarray}
\nonumber m_{p^+} &=& \left ( m_0 + {\sqrt3\over2} F - { 1 \over 2 \sqrt3 } D + {1\over2} f + {1\over2} d + EM_{p^+} \right) \bar p^+ p^+ \, ,
\\ \nonumber m_{n^0} &=& \left ( m_0 + {\sqrt3\over2} F - { 1 \over 2 \sqrt3 } D - {1\over2} f - {1\over2} d + EM_{n^0} \right) \bar n^0 n^0 \, ,
\\ \nonumber m_{\Sigma^+} &=& \left ( m_0 + { 1 \over \sqrt3 } D + f + EM_{\Sigma^+} \right) \bar \Sigma^+ \Sigma^+ \, ,
\\ m_{\Sigma^0} &=& \left ( m_0 + { 1 \over \sqrt3 } D + EM_{\Sigma^0} \right) \bar \Sigma^0 \Sigma^0 \, ,
\label{eq:diagonal}
\\ \nonumber m_{\Sigma^-} &=& \left ( m_0 + { 1 \over \sqrt3 } D - f + EM_{\Sigma^-} \right) \bar \Sigma^- \Sigma^- \, ,
\\ \nonumber m_{\Xi^0} &=& \left ( m_0 - {\sqrt3\over2} F - { 1 \over 2 \sqrt3 } D + {1\over2} f - {1\over2} d + EM_{\Xi^0} \right) \bar \Xi^0 \Xi^0 \, ,
\\ \nonumber m_{\Xi^-} &=& \left ( m_0 - {\sqrt3\over2} F - { 1 \over 2 \sqrt3 } D - {1\over2} f + {1\over2} d + EM_{\Xi^-} \right) \bar \Xi^- \Xi^- \, ,
\\ \nonumber m_{\Lambda^0} &=& \left ( m_0 - { 1 \over \sqrt3 } D + EM_{\Lambda^0} \right) \bar \Lambda^0 \Lambda^0 \, ,
\end{eqnarray}
where
\begin{eqnarray}
m_0 = m_{bare} + g_2 \langle \sigma_0 \rangle \, , D \equiv g_3 \langle \sigma_8 \rangle \, , F \equiv g_4 \langle \sigma_8 \rangle \, , d \equiv g_3 \langle \sigma_3 \rangle \, , f \equiv g_4 \langle \sigma_3 \rangle \, .
\end{eqnarray}
We can easily verify that these diagonal hadronic parts satisfy the well-known Gell-Mann-Okubo mass formula $2 \left( M_N + M_{\Xi} \right) = 3M_\Lambda +M_\Sigma$~\cite{Gell,Okubo}.

The off-diagonal terms are:
\begin{eqnarray}
\nonumber m_{\Sigma^0\Lambda^0} &=& { 1 \over \sqrt3 } d \Big ( \bar \Sigma^0 \Lambda^0 + \bar \Lambda^0 \Sigma^0 \Big ) \, ,
\\ \nonumber m_{\Sigma^{*0} \Lambda^0} &=& - { 1 \over \sqrt2 } m_{N\Delta} \Big ( \bar \Sigma^{*0} \Lambda^0 + \bar \Lambda^0 \Sigma^{*0} \Big ) \, ,
\\ \nonumber m_{p^+\Delta^{+}} &=& \sqrt{ 2 \over 3 } m_{N\Delta} \Big ( \bar p^+ \Delta^{+} + \bar \Delta^{+} p^+ \Big ) \, ,
\\ m_{n^0\Delta^{0}} &=& \sqrt{ 2 \over 3 } m_{N\Delta} \Big ( \bar n^0 \Delta^{0} + \bar \Delta^{0} n^0 \Big ) \, ,
\\ \nonumber m_{\Sigma^+\Sigma^{*+}} &=& \left ( { 1 \over \sqrt2 } M_{N\Delta} + { 1 \over \sqrt6 } m_{N\Delta} \right ) \Big ( \bar \Sigma^+ \Sigma^{*+} + \bar \Sigma^{*+} \Sigma^+ \Big ) \, ,
\\ \nonumber m_{\Sigma^0\Sigma^{*0}} &=& { 1 \over \sqrt2 } M_{N\Delta} \Big ( \bar \Sigma^0 \Sigma^{*0} + \bar \Sigma^{*0} \Sigma^0 \Big ) \, ,
\\ \nonumber m_{\Sigma^-\Sigma^{*-}} &=& \left ( { 1 \over \sqrt2 } M_{N\Delta} - { 1 \over \sqrt6 } m_{N\Delta} \right ) \Big ( \bar \Sigma^- \Sigma^{*-} + \bar \Sigma^{*-} \Sigma^- \Big ) \, ,
\\ \nonumber m_{\Xi^0\Xi^{*0}} &=& \left ( - { 1 \over \sqrt2 } M_{N\Delta} - { 1 \over \sqrt6 } m_{N\Delta} \right ) \Big ( \bar \Xi^0 \Xi^{*0} + \bar \Xi^{*0} \Xi^0 \Big ) \, ,
\\ \nonumber m_{\Xi^-\Xi^{*-}} &=& \left ( - { 1 \over \sqrt2 } M_{N\Delta} + { 1 \over \sqrt6 } m_{N\Delta} \right ) \Big ( \bar \Xi^- \Xi^{*-} + \bar \Xi^{*-} \Xi^- \Big ) \, ,
\end{eqnarray}
where
\begin{eqnarray}
M_{N\Delta} \equiv g_8 \langle \sigma_8 \rangle \, , m_{N\Delta} \equiv g_8 \langle \sigma_3 \rangle \, .
\end{eqnarray}

We use $m_{\Delta}$ to denote the decuplet baryon masses, which are all assumed to be 2 GeV:
\begin{eqnarray}
m_{\Delta} = 2~{\rm GeV} \, .
\label{eq:mdecuplet}
\end{eqnarray}

Summarizing all these diagonal and off-diagonal terms, we arrive at the following mass formulae:
\begin{eqnarray}
M_{p^+} &\in& \mbox{ eigenvalues of }
\left ( \begin{array}{cc}
m_0 + {\sqrt3\over2} F - { 1 \over 2 \sqrt3 } D + {1\over2} f + {1\over2} d + EM_{p^+} & \sqrt{ 2 \over 3 } m_{N\Delta}
\label{eq:proton}
\\ \sqrt{ 2 \over 3 } m_{N\Delta} & m_{\Delta}
\end{array} \right) \, ,
\\ M_{n^0} &\in& \mbox{ eigenvalues of }
\left ( \begin{array}{cc}
m_0 + {\sqrt3\over2} F - { 1 \over 2 \sqrt3 } D - {1\over2} f - {1\over2} d + EM_{n^0} & \sqrt{ 2 \over 3 } m_{N\Delta}
\\ \sqrt{ 2 \over 3 } m_{N\Delta} & m_{\Delta}
\end{array} \right) \, ,
\\ M_{\Sigma^+} &\in& \mbox{ eigenvalues of }
\left ( \begin{array}{cc}
m_0 + { 1 \over \sqrt3 } D + f + EM_{\Sigma^+} & { 1 \over \sqrt2 } M_{N\Delta} + { 1 \over \sqrt6 } m_{N\Delta}
\label{eq:sigmap}
\\ { 1 \over \sqrt2 } M_{N\Delta} + { 1 \over \sqrt6 } m_{N\Delta} & m_{\Delta}
\end{array} \right) \, ,
\\ M_{\Sigma^0}\&M_{\Lambda^0} &\in& \mbox{ eigenvalues of }
\left ( \begin{array}{ccc}
m_0 + { 1 \over \sqrt3 } D + EM_{\Sigma^0} & { 1 \over \sqrt3 } d & { 1 \over \sqrt2 } M_{N\Delta}
\\ { 1 \over \sqrt3 } d & m_0 - { 1 \over \sqrt3 } D + EM_{\Lambda^0} & - { 1 \over \sqrt2 } m_{N\Delta}
\\ { 1 \over \sqrt2 } M_{N\Delta} & - { 1 \over \sqrt2 } m_{N\Delta} & m_{\Delta}
\end{array} \right) \, ,
\\ M_{\Sigma^-} &\in& \mbox{ eigenvalues of }
\left ( \begin{array}{cc}
m_0 + { 1 \over \sqrt3 } D - f + EM_{\Sigma^-} & { 1 \over \sqrt2 } M_{N\Delta} - { 1 \over \sqrt6 } m_{N\Delta}
\\ { 1 \over \sqrt2 } M_{N\Delta} - { 1 \over \sqrt6 } m_{N\Delta} & m_{\Delta}
\end{array} \right) \, ,
\\ M_{\Xi^0} &\in& \mbox{ eigenvalues of }
\left ( \begin{array}{cc}
m_0 - {\sqrt3\over2} F - { 1 \over 2 \sqrt3 } D + {1\over2} f - {1\over2} d + EM_{\Xi^0} & - { 1 \over \sqrt2 } M_{N\Delta} - { 1 \over \sqrt6 } m_{N\Delta}
\label{eq:xi0}
\\ - { 1 \over \sqrt2 } M_{N\Delta} - { 1 \over \sqrt6 } m_{N\Delta} & m_{\Delta}
\end{array} \right) \, ,
\\ M_{\Xi^-} &\in& \mbox{ eigenvalues of }
\left ( \begin{array}{cc}
m_0 - {\sqrt3\over2} F - { 1 \over 2 \sqrt3 } D - {1\over2} f + {1\over2} d + EM_{\Xi^-} & - { 1 \over \sqrt2 } M_{N\Delta} + { 1 \over \sqrt6 } m_{N\Delta}
\\ - { 1 \over \sqrt2 } M_{N\Delta} + { 1 \over \sqrt6 } m_{N\Delta} & m_{\Delta}
\end{array} \right) \, .
\label{eq:xin}
\end{eqnarray}
In these equations there are altogether nine parameters: $m_0$, $D$, $F$, $d$, $f$, $m_{\Delta}$, $M_{N\Delta}$, $m_{N\Delta}$ and $EM_A$ ($EM_B$) with two constraints
\begin{eqnarray}
{d \over D} = { f \over F} = {m_{N\Delta} \over M_{N\Delta}} = {\langle \sigma_3 \rangle \over \langle \sigma_8 \rangle} \, ,
\end{eqnarray}
and so there are altogether seven free parameters. Moreover, in Sec.~\ref{sec:decupletmass} we shall find that we can always achieve a good fitting no matter which value of $m_{\Delta}$ is chosen, and so there are only six free parameters involved in the mass fitting. In the next section we shall use these equations to fit the eight lowest-lying flavor-octet baryon masses whose measurements are very accurate nowadays~\cite{Beringer:1900zz}:
\begin{eqnarray}
\nonumber && M_{p^+}^{phy} = 938.27 {\rm MeV} \, , M_{n^0}^{phy} = 939.57  {\rm MeV} \, ,
\\ && M_{\Sigma^+}^{phy} = 1189.37 {\rm MeV} \, , M_{\Sigma^0}^{phy} = 1192.64 {\rm MeV} \, , M_{\Sigma^-}^{phy} = 1197.45 {\rm MeV} \, ,
\\ \nonumber && M_{\Xi^0}^{phy} = 1314.86 {\rm MeV} \, , M_{\Xi^-}^{phy} = 1321.71 {\rm MeV} \, ,
\\ \nonumber && M_{\Lambda^0}^{phy} = 1115.68 {\rm MeV} \, .
\end{eqnarray}

\section{Numerical Analyses}
\label{sec:numerical}

In this section we perform numerical analyses and fit the eight lowest-lying flavor-octet baryon masses using the equations obtained in the previous section.
To do the fitting, we use the following variance equation:
\begin{eqnarray}
&& {\rm var} = Minimize \left (\sqrt { {1\over8} \sum_{N} {(M_N - M_N^{phy})^2 \over 1 {\rm~MeV}^2} } \right) \, ,
\\ \nonumber && \mbox{constrained by}
\left\{ \begin{array}{l}
{d \over D} = {f \over F} = {m_{N\Delta} \over M_{N\Delta}}
\\ m_{\Delta} = 2~{\rm GeV}
\end{array} \right. \, ,
\label{eq:chi}
\end{eqnarray}
where $M_N^{phy}$ are the experimental values of the eight lowest-lying flavor-octet baryon masses, and $M_N$ are our fitted masses calculated using Eqs.~(\ref{eq:proton})-(\ref{eq:xin}).

Our fitting is separated into four steps. The first step will be done in Sec.~\ref{sec:step1} where we only consider the $SU(3)$ flavor symmetry breaking, and $m_0$, $D$ and $F$ are assumed to be nonzero; the second step will be done in Sec.~\ref{sec:step2} where we do not consider the contributions of electromagnetic effects and decuplet baryons, and $m_0$, $D$, $F$, $d$ and $f$ are assumed to be nonzero; the third step will be done in Sec.~\ref{sec:step3} where the contributions of electromagnetic effects are included while the contributions of decuplet baryons are not included; the fourth step will be done in Sec.~\ref{sec:step4} where the contributions of decuplet baryons are included while the contributions of electromagnetic effects are not included; the fifth step will be done in Sec.~\ref{sec:step5} where all the nine parameters are assumed to be nonzero. In Sec.~\ref{sec:decupletmass} we shall discuss the dependence of our fittings on the decuplet baryon mass $m_{\Delta}$. In Appendix.~\ref{sec:equivalent} we shall further discuss the contributions of the current quark masses $m_{u,d,s}$.

\subsection{Step 1: $m_0$, $D$ and $F$}
\label{sec:step1}

In this subsection we only consider the $SU(3)$ flavor symmetry breaking, and $m_0$, $D$ and $F$ are assumed to be nonzero. In this case there are three free parameters involved. To do the fitting we use the following variance equation:
\begin{eqnarray}
&& {\rm var}_1 = Minimize \left (\sqrt { {1\over8} \sum_{N} {(M_N - M_N^{phy})^2 \over 1 {\rm~MeV}^2} } \right) \, ,
\\ \nonumber && \mbox{constrained by}
\left\{ \begin{array}{l}
d = f = M_{N\Delta} = m_{N\Delta} = EM_A = EM_B = 0
\\ m_{\Delta} = 2~{\rm GeV}
\end{array} \right. \, .
\label{eq:step1}
\end{eqnarray}
Very quickly we can obtain that ${\rm var}_1 = 3.66$, when $m_0 = 1151.19$ MeV, $D= 71.56$ MeV and $F=-219.03$ MeV. The results are shown in Table.~\ref{tab:step1}. In this step the difference between every fitted mass and its experimental value is less than 5.8 MeV, i.e., $| M_N - M_N^{phy} | < 5.8$ MeV. The maximum is due to the $\Lambda$ baryon that $| M_\Lambda - M_\Lambda^{phy} | = 5.8$ MeV.
\begin{table}[tbh]
\begin{center}
\caption{Step 1: $m_0$, $D$ and $F$ are assumed to be nonzero.}
\begin{tabular}{c|c}
\hline \hline
\multicolumn{2}{c}{Step 1, ${\rm var}_1 = 3.66$}
\\ \hline \hline
Parameters (MeV) & Masses (MeV)
\\ \hline
$\begin{array}{l}
m_0 = 1151.19
\\ D = 71.56
\\ F = -219.03
\\ d = 0
\\ f = 0
\\ M_{N\Delta} = 0
\\ m_{N\Delta} = 0
\\ EM_A = EM_B = 0
\end{array}$
&
$\begin{array}{l}
M_{p^+} = 940.85
\\ M_{n^0} = 940.85
\\ M_{\Sigma^+} = 1192.51
\\ M_{\Sigma^0} = 1192.51
\\ M_{\Sigma^-} = 1192.51
\\ M_{\Xi^0} = 1320.22
\\ M_{\Xi^-} = 1320.22
\\ M_{\Lambda^0} = 1109.88
\\ m_{\Delta} = 2000
\end{array}$
\\ \hline \hline
\end{tabular}
\label{tab:step1}
\end{center}
\end{table}

\subsection{Step 2: $m_0$, $D$, $F$, $d$ and $f$}
\label{sec:step2}

In this subsection we do not consider the contributions of electromagnetic effects and decuplet baryons, and $m_0$, $D$, $F$, $d$ and $f$ are assumed to be nonzero. There is one constraint among them and so there are four free parameters. To do the fitting we use the following variance equation:
\begin{eqnarray}
&& {\rm var}_2 = Minimize \left (\sqrt { {1\over8} \sum_{N} {(M_N - M_N^{phy})^2 \over 1 {\rm~MeV}^2} } \right) \, ,
\\ \nonumber && \mbox{constrained by}
\left\{ \begin{array}{l}
{d \over D} = {f \over F}
\\ M_{N\Delta} = m_{N\Delta} = EM_A = EM_B = 0
\\ m_{\Delta} = 2~{\rm GeV}
\end{array} \right. \, .
\label{eq:step2}
\end{eqnarray}
The results are shown in Table.~\ref{tab:step2}, where ${\rm var}_2 = 2.56$. Although ${\rm var}_2 < {\rm var}_1$, in this step the maximum difference between every fitted mass and its experimental value is still 5.8 MeV, i.e., $| M_N - M_N^{phy} | < 5.8$ MeV. This is because that the mass of the $\Lambda$ baryon is still not well fitted that $| M_\Lambda - M_\Lambda^{phy} | = 5.8$ MeV.
\begin{table}[tbh]
\begin{center}
\caption{Step 2: $m_0$, $D$, $F$, $d$ and $f$ are assumed to be nonzero.}
\begin{tabular}{c|c}
\hline \hline
\multicolumn{2}{c}{Step 2, ${\rm var}_2 = 2.56$}
\\ \hline \hline
Parameters (MeV) & Masses (MeV)
\\ \hline
$\begin{array}{l}
m_0 = 1151.19
\\ D = 71.57
\\ F = -219.02
\\ d = 1.37
\\ f = -4.20
\\ M_{N\Delta} = 0
\\ m_{N\Delta} = 0
\\ EM_A = EM_B = 0
\end{array}$
&
$\begin{array}{l}
M_{p^+} = 939.44
\\ M_{n^0} = 942.27
\\ M_{\Sigma^+} = 1188.32
\\ M_{\Sigma^0} = 1192.52
\\ M_{\Sigma^-} = 1196.71
\\ M_{\Xi^0} = 1317.43
\\ M_{\Xi^-} = 1323.00
\\ M_{\Lambda^0} = 1109.87
\\ m_{\Delta} = 2000
\end{array}$
\\ \hline \hline
\end{tabular}
\label{tab:step2}
\end{center}
\end{table}

\subsection{Step 3: $m_0$, $D$, $F$, $d$, $f$ and $EM_A$ ($EM_B$)}
\label{sec:step3}

In this subsection we include the contributions of electromagnetic effects, and $m_0$, $D$, $F$, $d$, $f$ and $EM_A$ ($EM_B$) are assumed to be nonzero. There is one constraint among them and so there are five free parameters. To do the fitting we use the following variance equation:
\begin{eqnarray}
&& {\rm var}_{\rm 3} = Minimize \left (\sqrt { {1\over8} \sum_{N} {(M_N - M_N^{phy})^2 \over 1 {\rm~MeV}^2} } \right) \, ,
\\ \nonumber && \mbox{constrained by}
\left\{ \begin{array}{l}
{d \over D} = {f \over F}
\\ M_{N\Delta} = m_{N\Delta} = 0
\\ m_{\Delta} = 2~{\rm GeV}
\end{array} \right. \, .
\label{eq:step3}
\end{eqnarray}
The results are shown in Table.~\ref{tab:step3}. We find that both the Scheme A and the Scheme B lead to similar results having ${\rm var}_{\rm 3} = 2.55$, and in both schemes the difference between every fitted mass and its experimental value is less than 5.7 MeV, i.e., $| M_N - M_N^{phy} | < 5.7$ MeV. Still the maximum is due to the $\Lambda$ baryon that $| M_\Lambda - M_\Lambda^{phy} | = 5.7$ MeV.
\begin{table}[tbh]
\begin{center}
\caption{Step 3: $m_0$, $D$, $F$, $d$, $f$ and $EM_A$ ($EM_B$) are assumed to be nonzero.}
\begin{tabular}{c|c||c|c}
\hline \hline
\multicolumn{2}{c||}{Step 3, Scheme A, ${\rm var}_{\rm 3} = 2.55$}
&
\multicolumn{2}{c}{Step 3, Scheme B, ${\rm var}_{\rm 3} = 2.55$}
\\ \hline \hline
Parameters (MeV) & Masses (MeV) & Parameters (MeV) & Masses (MeV)
\\ \hline
$\begin{array}{l}
m_0 = 1151.38
\\ D = 71.71
\\ F = -219.02
\\ d = 1.39
\\ f = -4.24
\\ M_{N\Delta} = 0
\\ m_{N\Delta} = 0
\\ EM_A = -0.38
\end{array}$
&
$\begin{array}{l}
M_{p^+} = 939.20
\\ M_{n^0} = 942.43
\\ M_{\Sigma^+} = 1188.17
\\ M_{\Sigma^0} = 1192.79
\\ M_{\Sigma^-} = 1196.64
\\ M_{\Xi^0} = 1317.55
\\ M_{\Xi^-} = 1322.80
\\ M_{\Lambda^0} = 1109.98
\\ m_{\Delta} = 2000
\end{array}$
&
$\begin{array}{l}
m_0 = 1151.11
\\ D = 71.76
\\ F = -219.12
\\ d = 1.45
\\ f = -4.42
\\ M_{N\Delta} = 0
\\ m_{N\Delta} = 0
\\ EM_B = -1.04
\end{array}$
&
$\begin{array}{l}
M_{p^+} = 939.14
\\ M_{n^0} = 942.46
\\ M_{\Sigma^+} = 1188.12
\\ M_{\Sigma^0} = 1192.89
\\ M_{\Sigma^-} = 1196.61
\\ M_{\Xi^0} = 1317.57
\\ M_{\Xi^-} = 1322.74
\\ M_{\Lambda^0} = 1110.02
\\ m_{\Delta} = 2000
\end{array}$
\\ \hline \hline
\end{tabular}
\label{tab:step3}
\end{center}
\end{table}

In this step the two electromagnetic parameters $EM_A$ and $EM_B$ are both negative, producing an electromagnetic mass difference between the proton and the neutron, $M^\gamma_{p-n} \equiv EM_{p^+} - EM_{n^0}<0$. This is not consistent with almost all theoretical calculations~\cite{Gasser:1982ap,WalkerLoud:2012bg,Shanahan:2012wa,Beane:2006fk,Blum:2010ym,Horsley:2012fw,deDivitiis:2013xla,Borsanyi:2013lga}, suggesting that this fitting is not well done.

\subsection{Step 4: $m_0$, $D$, $F$, $d$, $f$, $m_{\Delta}$, $M_{N\Delta}$ and $m_{N\Delta}$}
\label{sec:step4}

In this subsection we include the contributions of decuplet baryons, and $m_0$, $D$, $F$, $d$, $f$, $m_{\Delta}$, $M_{N\Delta}$ and $m_{N\Delta}$ are assumed to be nonzero. There are two constraints among them and so there are six free parameters (including $m_{\Delta}$). To do the fitting we fix $m_{\Delta} = 2~{\rm GeV}$ and use the following variance equation:
\begin{eqnarray}
&& {\rm var}_{\rm 4} = Minimize \left (\sqrt { {1\over8} \sum_{N} {(M_N - M_N^{phy})^2 \over 1 {\rm~MeV}^2} } \right) \, ,
\\ \nonumber && \mbox{constrained by}
\left\{ \begin{array}{l}
{d \over D} = {f \over F} = {m_{N\Delta} \over M_{N\Delta}}
\\ EM_A = EM_B = 0
\\ m_{\Delta} = 2~{\rm GeV}
\end{array} \right. \, .
\label{eq:step3B}
\end{eqnarray}
The results are shown in Table.~\ref{tab:step4}, where ${\rm var}_{\rm 4} = 0.44$. This is significantly smaller than ${\rm var}_1$ and ${\rm var}_2$. In this step the difference between every fitted mass and its experimental value is less than 0.61 MeV, i.e., $| M_N - M_N^{phy} | < 0.61$ MeV. This happens to the proton that $| M_{p^+} - M_{p^+}^{phy} | = 0.61$ MeV, and now the $\Lambda$ baryon is well fitted that $| M_\Lambda - M_\Lambda^{phy} | = 0.03$ MeV.
\begin{table}[tbh]
\begin{center}
\caption{Step 4: $m_0$, $D$, $F$, $d$, $f$, $m_{\Delta}$, $M_{N\Delta}$ and $m_{N\Delta}$ are assumed to be nonzero.}
\begin{tabular}{c|c}
\hline \hline
\multicolumn{2}{c}{Step 4, ${\rm var}_{\rm 4} = 0.44$}
\\ \hline \hline
Parameters (MeV) & Masses (MeV)
\\ \hline
$\begin{array}{l}
m_0 = 1163.95
\\ D = 83.52
\\ F = -232.01
\\ d = 1.40
\\ f = -3.90
\\ M_{N\Delta} = -175.16
\\ m_{N\Delta} = -2.94
\\ EM_A = EM_B = 0
\end{array}$
&
$\begin{array}{l}
M_{p^+} = 937.66
\\ M_{n^0} = 940.16
\\ M_{\Sigma^+} = 1188.99
\\ M_{\Sigma^0} = 1193.17
\\ M_{\Sigma^-} = 1197.33
\\ M_{\Xi^0} = 1315.28
\\ M_{\Xi^-} = 1321.26
\\ M_{\Lambda^0} = 1115.71
\\ m_{\Delta} = 2000
\end{array}$
\\ \hline \hline
\end{tabular}
\label{tab:step4}
\end{center}
\end{table}

\subsection{Step 5: $m_0$, $D$, $F$, $d$, $f$, $m_{\Delta}$, $M_{N\Delta}$, $m_{N\Delta}$ and $EM_A$ ($EM_B$)}
\label{sec:step5}

In this subsection we include all the nine parameters, $m_0$, $D$, $F$, $d$, $f$, $m_{\Delta}$, $M_{N\Delta}$, $m_{N\Delta}$ and $EM_A$ ($EM_B$), and assume that they are nonzero. There are two constraints among them and so there are seven free parameters (including $m_{\Delta}$). To do the fitting we fix $m_{\Delta} = 2~{\rm GeV}$ and use the following variance equation:
\begin{eqnarray}
&& {\rm var}_5 = Minimize \left (\sqrt { {1\over8} \sum_{N} {(M_N - M_N^{phy})^2 \over 1 {\rm~MeV}^2} } \right) \, ,
\\ \nonumber && \mbox{constrained by}
\left\{ \begin{array}{l}
{d \over D} = {f \over F} = {m_{N\Delta} \over M_{N\Delta}}
\\ m_{\Delta} = 2~{\rm GeV}
\end{array} \right. \, .
\label{eq:step5}
\end{eqnarray}
The results are shown in Table.~\ref{tab:step5}. When using the Scheme A, we obtain ${\rm var}_5 = 0.06$, and in this case the difference between every fitted mass and its experimental value is less than 0.11 MeV, i.e., $| M_N - M_N^{phy} | < 0.11$ MeV. When using the Scheme B, we obtain ${\rm var}_5 = 0.10$, and in this case the difference between every fitted mass and its experimental value is less than 0.16 MeV, i.e., $| M_N - M_N^{phy} | < 0.16$ MeV. Hence, the Scheme A seems slightly better than the Scheme B.

We can solve the eigenvectors of Eqs.~(\ref{eq:proton}), (\ref{eq:sigmap}) and (\ref{eq:xi0}) and estimate the decuplet components contained in the $N$, $\Sigma$ and $\Xi$ baryons, which are around $5\times10^{-6}$, 2.4\% and 3.4\%, respectively, using both schemes. We can also estimate the $\Sigma^0-\Lambda^0$ mixing~\cite{Zhu:1998ai,Muller:1999vd,Yagisawa:2001gz,Shinmura:2002te,Sasaki:2003av}, which is at the level of $10^{-4}$:
\begin{eqnarray}
| \Lambda^0 \rangle_{phy} = 0.9999 | \Lambda^0 \rangle - 0.0137 | \Sigma^0 \rangle -  0.0043 | \Sigma^{*0} \rangle \, .
\end{eqnarray}
\begin{table}[tbh]
\begin{center}
\caption{Step 5: all the nine parameters are assumed to be nonzero.}
\begin{tabular}{c|c||c|c}
\hline \hline
\multicolumn{2}{c||}{Step 5, Scheme A, ${\rm var}_5 = 0.06$}
&
\multicolumn{2}{c}{Step 5, Scheme B, ${\rm var}_5 = 0.10$}
\\ \hline \hline
Parameters (MeV) & Masses (MeV) & Parameters (MeV) & Masses (MeV)
\\ \hline
$\begin{array}{l}
m_0 = 1164.03
\\ D = 83.70
\\ F = -232.59
\\ d = 1.36
\\ f = -3.78
\\ M_{N\Delta} = -179.02
\\ m_{N\Delta} = -2.91
\\ EM_A = 0.98
\end{array}$
&
$\begin{array}{l}
M_{p^+} = 938.20
\\ M_{n^0} = 939.64
\\ M_{\Sigma^+} = 1189.42
\\ M_{\Sigma^0} = 1192.53
\\ M_{\Sigma^-} = 1197.52
\\ M_{\Xi^0} = 1314.90
\\ M_{\Xi^-} = 1321.67
\\ M_{\Lambda^0} = 1115.68
\\ m_{\Delta} = 2000
\end{array}$
&
$\begin{array}{l}
m_0 = 1164.74
\\ D = 83.73
\\ F = -232.44
\\ d = 1.25
\\ f = -3.47
\\ M_{N\Delta} = -179.34
\\ m_{N\Delta} = -2.68
\\ EM_B = 2.10
\end{array}$
&
$\begin{array}{l}
M_{p^+} = 938.16
\\ M_{n^0} = 939.68
\\ M_{\Sigma^+} = 1189.43
\\ M_{\Sigma^0} = 1192.48
\\ M_{\Sigma^-} = 1197.55
\\ M_{\Xi^0} = 1314.93
\\ M_{\Xi^-} = 1321.63
\\ M_{\Lambda^0} = 1115.69
\\ m_{\Delta} = 2000
\end{array}$
\\ \hline \hline
\end{tabular}
\label{tab:step5}
\end{center}
\end{table}

The two assumptions to calculate electromagnetic terms, the Scheme A and the Scheme B, give different $M^\gamma_{p-n} \equiv EM_{p^+} - EM_{n^0}$. This suggests that to well separate hadronic and electromagnetic effects, one needs more reasonable (reliable) calculations. The Scheme A produces $M^\gamma_{p-n} = 0.98$ MeV, $M^\gamma_{\Sigma^+-\Sigma^-} = 0$, $M^\gamma_{\Xi^0-\Xi^-} = - 0.98$ MeV, and $M^\gamma_{\Sigma^++\Sigma^--2\Sigma^0} = 1.96$ MeV. These results are just consistent with the data estimated in Ref.~\cite{Gasser:1982ap} that $M^\gamma_{p-n} = EM_A = 0.76\pm0.30$ MeV, $M^\gamma_{\Sigma^+-\Sigma^-} = -0.17\pm0.30$, $M^\gamma_{\Xi^0-\Xi^-} = - 0.86\pm0.30$ MeV, and $M^\gamma_{\Sigma^++\Sigma^--2\Sigma^0} = 1.78\pm0.14$. The Scheme B produces $M^\gamma_{p-n} = 0.70$ MeV, $M^\gamma_{\Sigma^+-\Sigma^-} = - 0.70$ MeV, $M^\gamma_{\Xi^0-\Xi^-} = - 1.40$ MeV, and $M^\gamma_{\Sigma^++\Sigma^--2\Sigma^0} = 1.40$ MeV. These results are not consistent with Ref.~\cite{Gasser:1982ap}. Again the Scheme A seems to be better than the Scheme B.

\subsection{Dependence on the decuplet baryon mass $m_{\Delta}$}
\label{sec:decupletmass}

\begin{figure*}[hbt]
\begin{center}
\scalebox{0.6}{\includegraphics{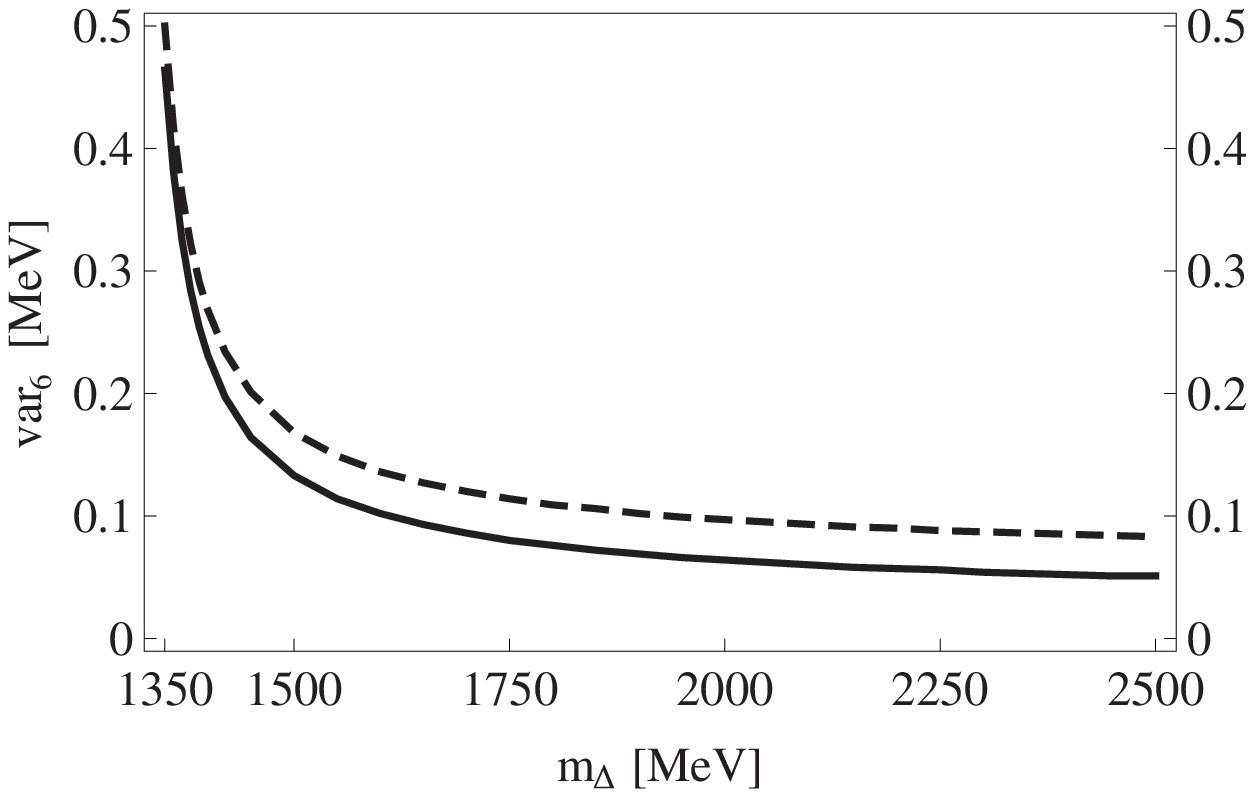}}
\scalebox{0.6}{\includegraphics{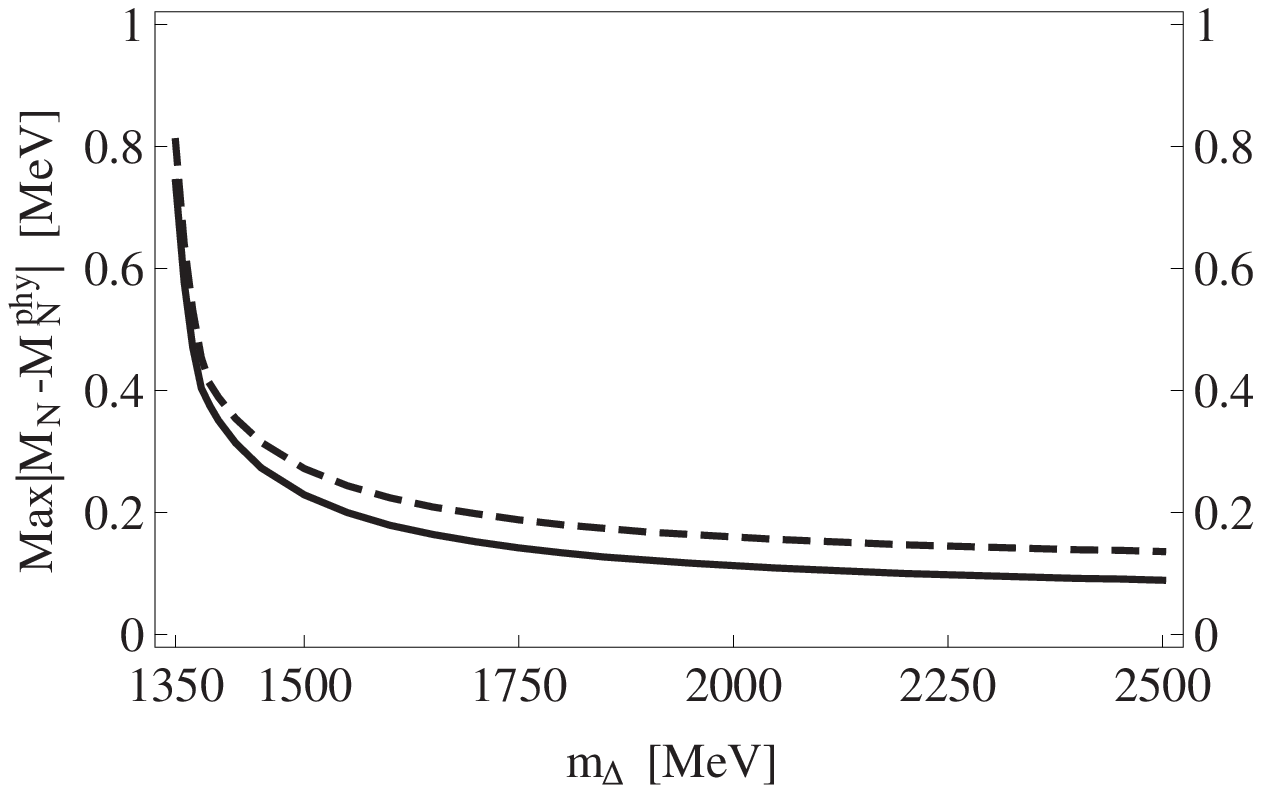}}
\caption{The variance ${\rm var}_6$ and the maximum difference between every fitted mass and its experimental value, ${\rm Max}|M_N - M_N^{phy}|$, as functions of the decuplet baryon mass $m_{\Delta}$. We use the solid and dashed curves to denote the results obtained using the Scheme A and the Scheme B, respectively.}
\label{fig:mixing}
\end{center}
\end{figure*}

\begin{figure*}[hbt]
\begin{center}
\scalebox{0.575}{\includegraphics{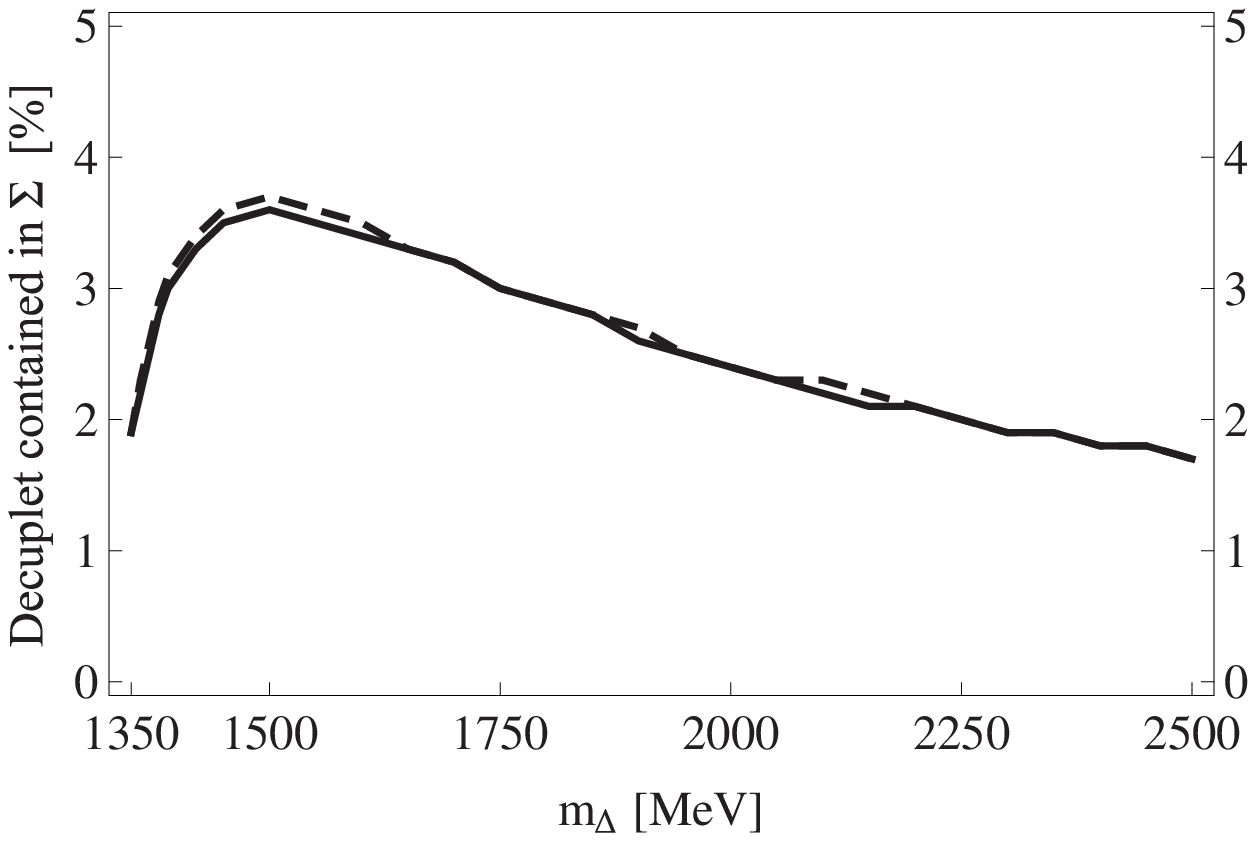}}
\scalebox{0.6}{\includegraphics{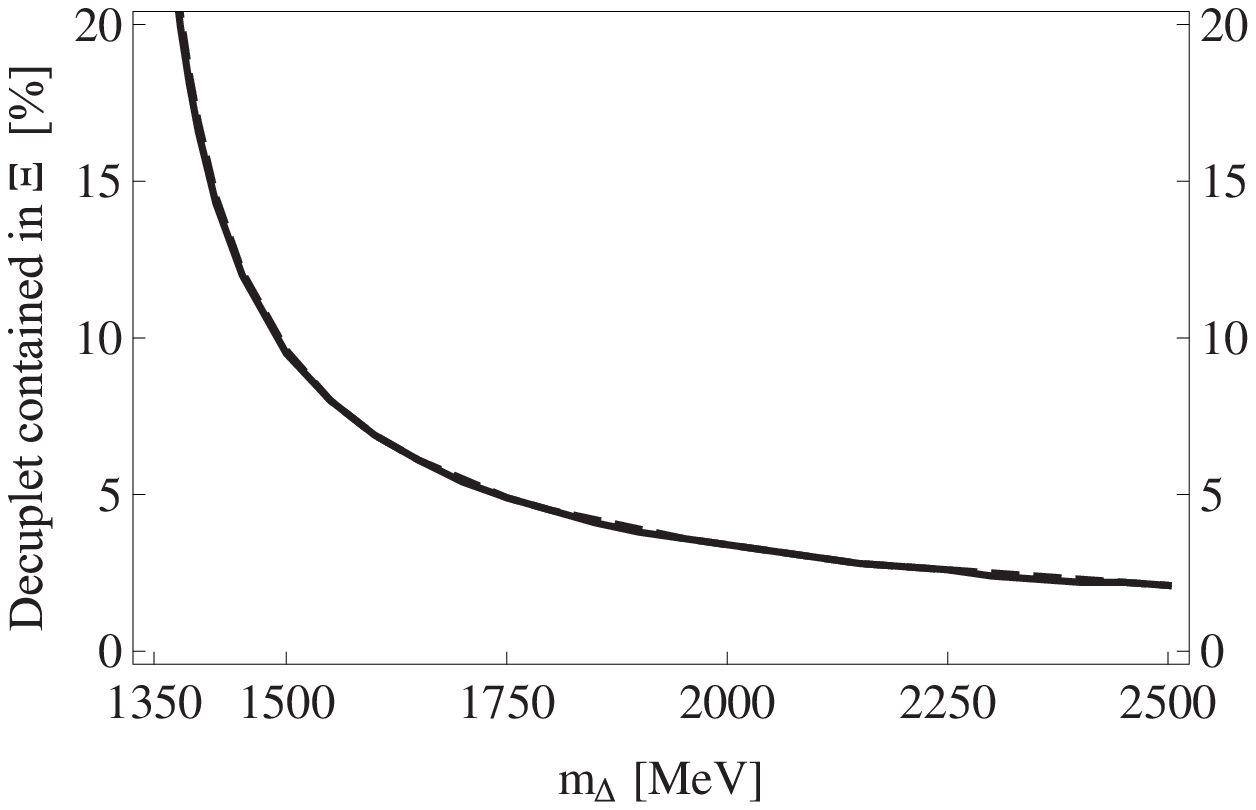}}
\caption{The decuplet components contained in the $\Sigma$ and $\Xi$ baryons, as functions of the decuplet baryon mass $m_{\Delta}$. We use the solid and dashed curves to denote the results obtained using the Scheme A and the Scheme B, respectively, while they are almost the same.}
\label{fig:decuplet}
\end{center}
\end{figure*}

In this subsection we study the dependence of our results on the decuplet baryon mass $m_{\Delta}$. We change $m_{\Delta}$ from 1350 MeV to 2500 MeV. To do the fitting we use the following variance equation:
\begin{eqnarray}
&& {\rm var}_6 = Minimize \left (\sqrt { {1\over8} \sum_{N} {(M_N - M_N^{phy})^2 \over 1 {\rm~MeV}^2} } \right) \, ,
\\ \nonumber && \mbox{constrained by}~~
{d \over D} = {f \over F} = {m_{N\Delta} \over M_{N\Delta}} \, ,
\\ \nonumber && \mbox{and } m_{\Delta} \mbox{ changing from 1800 MeV to 2500 MeV} \, .
\label{eq:step5}
\end{eqnarray}
We show ${\rm var}_6$ and ${\rm Max}|M_N - M_N^{phy}|$ (the maximum difference between every fitted mass and its experimental value) as functions of $m_{\Delta}$ in Fig.~\ref{fig:mixing}. We clearly see that we can always achieve a good fitting no matter which value of $m_{\Delta}$ is chosen. Hence, we have proved that there are only six free parameters involved in our fittings. However, the mixing parameters $M_{N\Delta}$ and $m_{N\Delta}$ do depend on $m_{\Delta}$, and so the decuplet components contained in the $\Sigma$ and $\Xi$ baryons also depend on $m_{\Delta}$. We show them in Fig.~\ref{fig:decuplet}.

Here we would like to note again that the Lagrangians containing $\pi^a$ can also contribute, as discussed in Sec.~\ref{sec:transform}. In this subsection we have found that we can use $m_{\Delta}$ as small as 1500 MeV to achieve a good fitting, which value is smaller than the total masses of the $J^P=0^-$ pseudoscalar mesons $\pi$ and the $J^P=(3/2)^+$ decuplet baryons $\Sigma(1385)$ and $\Xi(1530)$. This may suggest that the contributions of the $J^P=(1/2)^+$ decuplet baryons, which are investigated in this paper, may be equivalently estimated using the Lagrangians containing these $J^P=0^-$ pseudoscalar mesons and $J^P=(3/2)^+$ decuplet baryons. However, these are beyond the scope of this paper, and so we did not investigate them.

\section{Summary}
\label{sec:summary}

We studied the isospin symmetry breaking and mass splittings of the lowest-lying flavor-octet baryons. We included three kinds of baryon mass terms: the bare mass term, the electromagnetic terms and the spontaneous chiral symmetry breaking terms. Particulary, we included the mixing term between flavor-octet and flavor-decuplet baryons. This assumes that the lowest-lying $\Sigma$ and $\Xi$ baryons contain a few decuplet components and so are not purely flavor-octet. Summarizing all these terms, we obtained the formulae to calculate octet baryon masses, and we used them to fit the eight lowest-lying flavor-octet baryon masses whose measurements are very accurate nowadays.

\begin{figure*}[hbt]
\begin{center}
\scalebox{0.7}{\includegraphics{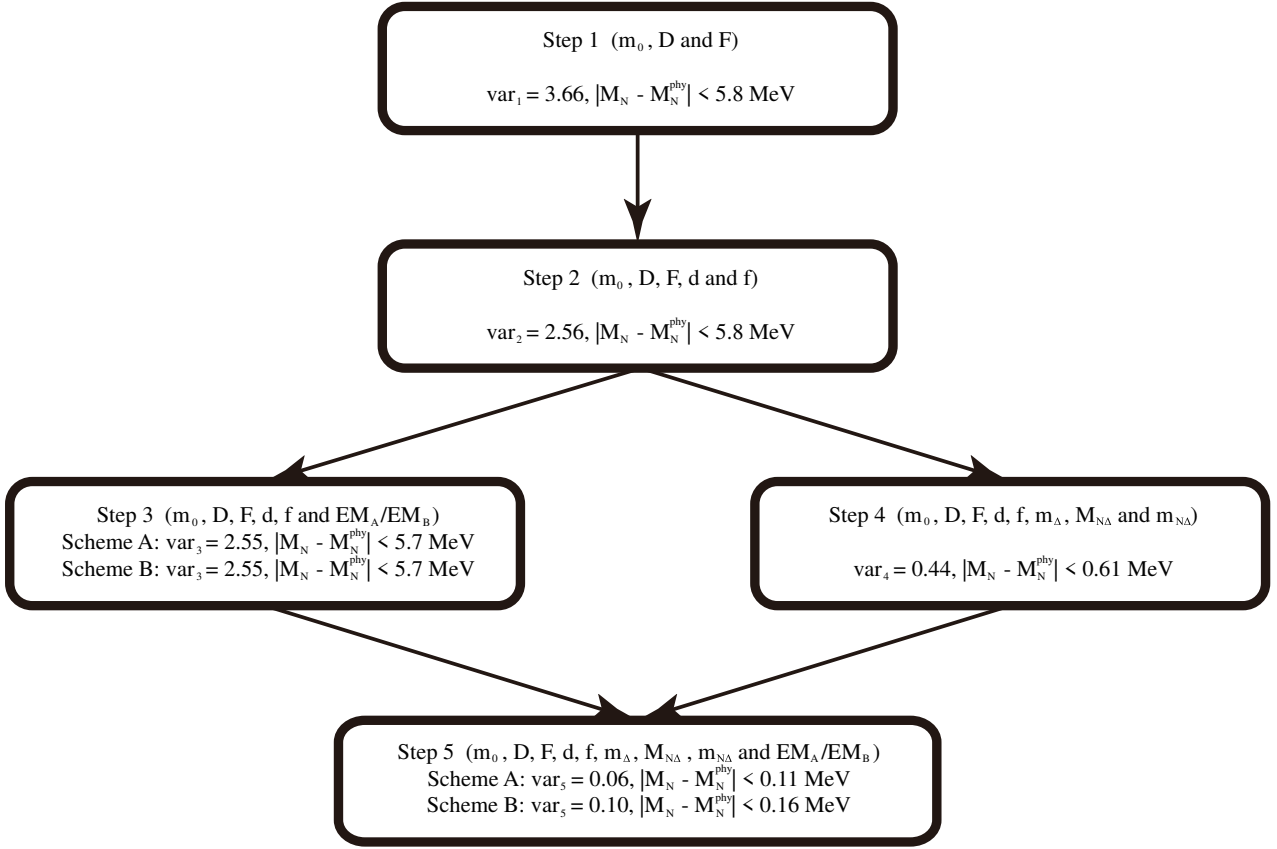}}
\caption{Our fittings are separated into five steps.}
\label{fig:procedure}
\end{center}
\end{figure*}

Our fittings was separated into five steps. We summarize all the results and show them in Fig.~\ref{fig:procedure}. In the last step we included all the nine parameters, $m_0$, $D$, $F$, $d$, $f$, $m_{\Delta}$, $M_{N\Delta}$, $m_{N\Delta}$ and $EM_A$ ($EM_B$). There are two constraints among these parameters, and so there are seven free parameters (including $m_{\Delta}$). After fixing $m_{\Delta}=2$ GeV, we obtain the results: when using the Scheme A, the difference between every fitted mass and its experimental value is less than 0.11 MeV, i.e., $| M_N - M_N^{phy} | < 0.11$ MeV; when using the Scheme B, this value is less than 0.16 MeV, i.e., $| M_N - M_N^{phy} | < 0.16$ MeV. Moreover, we have shown that a good fitting can be always achieved no matter which value of $m_{\Delta}$ is chosen, and so there are six free parameters involved in the mass fitting.

We estimated the decuplet components contained in the $\Sigma$ and $\Xi$ baryons, which are around 2.4\% and 3.4\%, respectively, when using $m_{\Delta}=2$ GeV. We also estimated the $\Sigma^0-\Lambda^0$ mixing, which is at the level of $10^{-4}$. To end this paper we would like to note again that the flavor-singlet and flavor-octet baryons can also mix through their common $[(\mathbf{\bar 3}, \mathbf{3})\oplus(\mathbf{3}, \mathbf{\bar 3})]$ chiral components. We have studied this mixing and the results are still in preparations~\cite{future}.

\section*{Acknowledgments}

This work is partly supported by the National Natural Science Foundation of China under Grant No. 11205011.

\appendix

\section{The Quark Mass Terms}
\label{sec:equivalent}

In this appendix we discuss the quark mass terms, $m_u$, $m_d$ and $m_s$. We shall prove that they are numerically equivalent to the spontaneous chiral symmetry breaking terms $g_2 \langle \sigma_0 \rangle \bar N^N N^N$ and $g_4 \langle \sigma_{3,8} \rangle ({\bf F}^{3,8}_{(8)})_{NM} \bar N^N N^M$, i.e., $m_{u,d,s}$ and $m_0$, $F$, $f$ are numerically equivalent. In the following we show how to relate these six parameters.

Assuming the flavor-octet baryon masses come from the following quark mass terms:
\begin{eqnarray}
&& m_{p^+} = 2 m_u + m_d \, , m_{n^0} = m_u + 2 m_d \, ,
\\ \nonumber && m_{\Sigma^+} = 2 m_u + m_s \, , m_{\Sigma^0} = m_u + m_d + m_s \, , m_{\Sigma^-} = 2 m_d + m_s \, ,
\\ \nonumber && m_{\Xi^0} = m_u + 2 m_s \, , m_{\Xi^-} = m_d + 2 m_s \, ,
\\ \nonumber && m_{\Lambda^0} = m_u + m_d + m_s \, .
\end{eqnarray}
Inserting
\begin{eqnarray}
m_u &=& {1\over3} m_0 + {1\over2\sqrt3}F + {1\over2} f \, ,
\label{eq:transform}
\\ \nonumber m_d &=& {1\over3} m_0 + {1\over2\sqrt3}F - {1\over2} f \, ,
\\ \nonumber m_s &=& {1\over3} m_0 - {1\over\sqrt3}F \, ,
\end{eqnarray}
these equations are changed to
\begin{eqnarray}
&& m_{p^+} = m_0 + {\sqrt3\over2} F + {1\over2} f \, , m_{n^0} = m_0 + {\sqrt3\over2} F - {1\over2} f \, ,
\\ \nonumber && m_{\Sigma^+} = m_0 + f \, , m_{\Sigma^0} = m_0 \, , m_{\Sigma^-} = m_0 - f \, ,
\\ \nonumber && m_{\Xi^0} = m_0 - {\sqrt3\over2} F + {1\over2} f \, , m_{\Xi^-} = m_0 - {\sqrt3\over2} F - {1\over2} f \, ,
\\ \nonumber && m_{\Lambda^0} = m_0 \, ,
\end{eqnarray}
which are just Eqs.~(\ref{eq:diagonal}) keeping only $m_0$, $F$ and $f$. Therefore, we have proved that $m_{u,d,s}$ and $m_0$, $F$, $f$ are numerically equivalent.


\begin{thebibliography}{99}

\bibitem{Feynman}
  R.~P.~Feynman and G.~Speisman,
  Phys.\ Rev. {\bf 94}, 500 (1954).

\bibitem{Gasser:1982ap}
  J.~Gasser and H.~Leutwyler,
  Phys.\ Rept.\  {\bf 87}, 77 (1982).

\bibitem{Shanahan:2012wa}
  P.~E.~Shanahan, A.~W.~Thomas and R.~D.~Young,
  Phys.\ Lett.\ B {\bf 718}, 1148 (2013).

\bibitem{Shanahan:2013xw}
  P.~E.~Shanahan, A.~W.~Thomas and R.~D.~Young,
  Phys.\ Rev.\ D {\bf 87}, 114515 (2013).

\bibitem{Praszalowicz:1992gn}
  M.~Praszalowicz, A.~Blotz and K.~Goeke,
  Phys.\ Rev.\ D {\bf 47}, 1127 (1993).

\bibitem{Yang:2010id}
  G.~-S.~Yang, H.~-C.~Kim and M.~V.~Polyakov,
  Phys.\ Lett.\ B {\bf 695}, 214 (2011).

\bibitem{WalkerLoud:2012bg}
  A.~Walker-Loud, C.~E.~Carlson and G.~A.~Miller,
  Phys.\ Rev.\ Lett.\  {\bf 108}, 232301 (2012).

\bibitem{Adami:1993xz}
  C.~Adami, E.~G.~Drukarev and B.~L.~Ioffe,
  Phys.\ Rev.\ D {\bf 48}, 2304 (1993)
  [Erratum-ibid.\ D {\bf 52}, 4254 (1995)].

\bibitem{Forkel:1996ty}
  H.~Forkel and M.~Nielsen,
  Phys.\ Rev.\ D {\bf 55}, 1471 (1997).

\bibitem{Henley:1996hr}
  E.~M.~Henley and T.~Meissner,
  Phys.\ Rev.\ C {\bf 55}, 3093 (1997).

\bibitem{Narison:2002pw}
  S.~Narison,
  Camb.\ Monogr.\ Part.\ Phys.\ Nucl.\ Phys.\ Cosmol.\  {\bf 17}, 1 (2002).

\bibitem{ioffe}
  B.~L.~Ioffe, V.~S.~Fadin, V.~A.~Khoze and L.~N.~Lipatov, {\it Quantum Chromodynamics: Perturbative and Nonperturbative Aspects}, Cambridge University Press, 2010.

\bibitem{dashen}
  R.~F.~Dashen,
  Phys.\ Rev.\  {\bf 183}, 1245 (1969).

\bibitem{Donoghue:1993hj}
  J.~F.~Donoghue, B.~R.~Holstein and D.~Wyler,
  Phys.\ Rev.\ D {\bf 47}, 2089 (1993).

\bibitem{Bijnens:1993ae}
  J.~Bijnens,
  Phys.\ Lett.\ B {\bf 306}, 343 (1993).

\bibitem{Li:1986iya}
  B.~-A.~Li, M.~-L.~Yan and K.~-F.~Liu,
  Phys.\ Lett.\ B {\bf 177}, 409 (1986).

\bibitem{Jain:1989kn}
  P.~Jain, R.~Johnson, N.~W.~Park, J.~Schechter and H.~Weigel,
  Phys.\ Rev.\ D {\bf 40}, 855 (1989).

\bibitem{Durr:2008zz}
  S.~Durr, Z.~Fodor, J.~Frison, C.~Hoelbling, R.~Hoffmann, S.~D.~Katz, S.~Krieg and T.~Kurth {\it et al.},
  Science {\bf 322}, 1224 (2008).

\bibitem{Duncan:2004ys}
  A.~Duncan, E.~Eichten and R.~Sedgewick,
  Phys.\ Rev.\ D {\bf 71}, 094509 (2005).

\bibitem{Basak:2013iw}
  S.~Basak {\it et al.}  [MILC Collaboration],
  PoS CD {\bf 12}, 030 (2013).

\bibitem{Bhattacharya:2013ehc}
  T.~Bhattacharya, S.~D.~Cohen, R.~Gupta, A.~Joseph and H.~-W.~Lin,
  arXiv:1306.5435 [hep-lat].

\bibitem{Green:2012ej}
  J.~R.~Green, J.~W.~Negele, A.~V.~Pochinsky, S.~N.~Syritsyn, M.~Engelhardt and S.~Krieg,
  Phys.\ Rev.\ D {\bf 86}, 114509 (2012).

\bibitem{Beane:2006fk}
  S.~R.~Beane, K.~Orginos and M.~J.~Savage,
  Nucl.\ Phys.\ B {\bf 768}, 38 (2007).

\bibitem{Horsley:2012fw}
  R.~Horsley {\it et al.}  [QCDSF and UKQCD Collaborations],
  Phys.\ Rev.\ D {\bf 86}, 114511 (2012).

\bibitem{Borsanyi:2013lga}
  S.~.Borsanyi, S.~D¨¹rr, Z.~Fodor, J.~Frison, C.~Hoelbling, S.~D.~Katz, S.~Krieg and T.~.Kurth {\it et al.},
  Phys.\ Rev.\ Lett.\  {\bf 111}, 252001 (2013).

\bibitem{deDivitiis:2013xla}
  G.~M.~de Divitiis, R.~Frezzotti, V.~Lubicz, G.~Martinelli, R.~Petronzio, G.~C.~Rossi, F.~Sanfilippo and S.~Simula {\it et al.},
  Phys.\ Rev.\ D {\bf 87}, 114505 (2013).

\bibitem{Blum:2010ym}
  T.~Blum, R.~Zhou, T.~Doi, M.~Hayakawa, T.~Izubuchi, S.~Uno and N.~Yamada,
  Phys.\ Rev.\ D {\bf 82}, 094508 (2010).

\bibitem{Bietenholz:2011qq}
  W.~Bietenholz, V.~Bornyakov, M.~Gockeler, R.~Horsley, W.~G.~Lockhart, Y.~Nakamura, H.~Perlt and D.~Pleiter {\it et al.},
  Phys.\ Rev.\ D {\bf 84}, 054509 (2011).

\bibitem{Aoki:2008sm}
  S.~Aoki {\it et al.}  [PACS-CS Collaboration],
  Phys.\ Rev.\ D {\bf 79}, 034503 (2009).

\bibitem{Fodor:2012gf}
  Z.~Fodor and C.~Hoelbling,
  Rev.\ Mod.\ Phys.\  {\bf 84}, 449 (2012).

\bibitem{Colangelo:2010et}
  G.~Colangelo, S.~Durr, A.~Juttner, L.~Lellouch, H.~Leutwyler, V.~Lubicz, S.~Necco and C.~T.~Sachrajda {\it et al.},
  Eur.\ Phys.\ J.\ C {\bf 71}, 1695 (2011).

\bibitem{Portelli:2013jla}
  A.~Portelli,
  PoS KAON {\bf 13}, 023 (2013).

\bibitem{Weinberg:1969hw}
  S.~Weinberg,
  Phys.\ Rev.\  {\bf 177}, 2604 (1969).

\bibitem{Weinberg:1990xn}
  S.~Weinberg,
  Phys.\ Rev.\ Lett.\  {\bf 65}, 1177 (1990).

\bibitem{Leinweber:1994nm}
  D.~B.~Leinweber,
  Phys.\ Rev.\  D {\bf 51}, 6383 (1995).

\bibitem{Ioffe:1981kw}
  B.~L.~Ioffe,
  Nucl.\ Phys.\  B {\bf 188}, 317 (1981)
  [Erratum-ibid.\  B {\bf 191}, 591 (1981)].

\bibitem{Chung:1981cc}
  Y.~Chung, H.~G.~Dosch, M.~Kremer and D.~Schall,
  Nucl.\ Phys.\  B {\bf 197} (1982) 55.

\bibitem{Espriu:1983hu}
  D.~Espriu, P.~Pascual and R.~Tarrach,
  Nucl.\ Phys.\  B {\bf 214} (1983) 285.

\bibitem{Cohen:1996sb}
  T.~D.~Cohen and X.~D.~Ji,
  Phys.\ Rev.\  D {\bf 55}, 6870 (1997).

\bibitem{Nagata:2007di}
  K.~Nagata, A.~Hosaka and V.~Dmitrasinovic,
  Mod.\ Phys.\ Lett.\  A {\bf 23}, 2381 (2008).

\bibitem{Cohen:2002st}
  T.~D.~Cohen and L.~Y.~Glozman,
  Int.\ J.\ Mod.\ Phys.\  A {\bf 17}, 1327 (2002).

\bibitem{Jido:2001nt}
  D.~Jido, M.~Oka and A.~Hosaka,
  Prog.\ Theor.\ Phys.\  {\bf 106}, 873 (2001).

\bibitem{Chen:2012ut}
  H.~-X.~Chen,
  Eur.\ Phys.\ J.\ C {\bf 72}, 2204 (2012).

\bibitem{Chen:2013gnu}
  H.~-X.~Chen,
  Adv.\ High Energy Phys.\  {\bf 2013}, 750591 (2013).

\bibitem{Chen:2013jra}
  H.~-X.~Chen,
  Eur.\  Phys.\  J.\ C {\bf 73}, 2628 (2013).

\bibitem{Chen:2008qv}
  H.~-X.~Chen, V.~Dmitrasinovic, A.~Hosaka, K.~Nagata and S.~-L.~Zhu,
  Phys.\ Rev.\ D {\bf 78}, 054021 (2008).

\bibitem{Dmitrasinovic:2011yf}
  V.~Dmitrasinovic and H.~-X.~Chen,
  Eur.\ Phys.\ J.\ C {\bf 71}, 1543 (2011).

\bibitem{Chen:2012vs}
  H.~-X.~Chen,
  Eur.\ Phys.\ J.\ C {\bf 72}, 2129 (2012).

\bibitem{Chen:2013efa}
  H.~-X.~Chen and V.~Dmitrasinovic,
  Phys.\ Rev.\ D {\bf 88}, 036013 (2013).

\bibitem{Beringer:1900zz}
  J.~Beringer {\it et al.}  [Particle Data Group Collaboration],
  Phys.\ Rev.\ D {\bf 86}, 010001 (2012).

\bibitem{Chen:2009sf}
  H.~-X.~Chen, V.~Dmitrasinovic and A.~Hosaka,
  Phys.\ Rev.\ D {\bf 81}, 054002 (2010).

\bibitem{Chen:2010ba}
  H.~X.~Chen, V.~Dmitrasinovic and A.~Hosaka,
  Phys.\ Rev.\  D {\bf 83}, 014015 (2011).

\bibitem{Chen:2011rh}
  H.~-X.~Chen, V.~Dmitrasinovic and A.~Hosaka,
  Phys.\ Rev.\ C {\bf 85}, 055205 (2012).

\bibitem{future}
  H.~-X.~Chen, {\it Isospin Symmetry Breaking and Octet Baryon Masses due to Their Mixing with the Singlet Baryon}, in preparations.

\bibitem{Weinberg:1978kz}
  S.~Weinberg,
  Physica A {\bf 96}, 327 (1979).

\bibitem{Gasser:1983yg}
  J.~Gasser and H.~Leutwyler,
  Annals Phys.\  {\bf 158}, 142 (1984).

\bibitem{Bernard:1995dp}
  V.~Bernard, N.~Kaiser and U.~-G.~Meissner,
  Int.\ J.\ Mod.\ Phys.\ E {\bf 4}, 193 (1995).

\bibitem{Pich:1995bw}
  A.~Pich,
  Rept.\ Prog.\ Phys.\  {\bf 58}, 563 (1995).

\bibitem{Ecker:1994gg}
  G.~Ecker,
  Prog.\ Part.\ Nucl.\ Phys.\  {\bf 35}, 1 (1995).

\bibitem{Scherer:2012xha}
  S.~Scherer and M.~R.~Schindler,
  Lect.\ Notes Phys.\  {\bf 830}, pp.1 (2012).

\bibitem{Pascalutsa:1998pw}
  V.~Pascalutsa,
  Phys.\ Rev.\ D {\bf 58}, 096002 (1998).

\bibitem{Pascalutsa:1999zz}
  V.~Pascalutsa and R.~Timmermans,
  Phys.\ Rev.\ C {\bf 60}, 042201 (1999).

\bibitem{Ren:2013oaa}
  X.~-L.~Ren, L.~-S.~Geng and J.~Meng,
  arXiv:1307.1896.

\bibitem{Coleman}
  S.~R.~Coleman and S.~L.~Glashow,
  Phys.\ Rev.\ Lett.\  {\bf 6}, 423 (1961).

\bibitem{Gell}
  M.~Gell-Mann,
  Phys.\ Rev.\  {\bf 125}, 1067 (1962).

\bibitem{Okubo}
  S.~Okubo,
  Prog.\ Theor.\ Phys.\  {\bf 27}, 949 (1962).

\bibitem{Zhu:1998ai}
  S.~-L.~Zhu, W.~Y.~P.~Hwang and Z.~-s.~Yang,
  Phys.\ Rev.\ D {\bf 57}, 1524 (1998).

\bibitem{Muller:1999vd}
  H.~Muller and J.~R.~Shepard,
  J.\ Phys.\ G {\bf 26}, 1049 (2000).

\bibitem{Yagisawa:2001gz}
  N.~Yagisawa, T.~Hatsuda and A.~Hayashigaki,
  Nucl.\ Phys.\ A {\bf 699}, 665 (2002).

\bibitem{Shinmura:2002te}
  S.~Shinmura, K.~S.~Myint, T.~Harada and Y.~Akaishi,
  J.\ Phys.\ G {\bf 28}, L1 (2002).

\bibitem{Sasaki:2003av}
  K.~Sasaki, T.~Inoue and M.~Oka,
  Few Body Syst.\ Suppl.\  {\bf 14}, 277 (2003).

\end{thebibliography}
\end{document}